\documentclass{aastex631}

\shorttitle{Hubble Ultraviolet Spectroscopy of Lucy Targets}
\shortauthors{Humes et al.}

\graphicspath{{./}{figures/}}

\begin{document}

\title{Ultraviolet Spectroscopy of Lucy Mission Targets with the Hubble Space Telescope}

\correspondingauthor{Oriel Humes}
\email{oah28@nau.edu}

\author[0000-0002-1700-5364]{Oriel A. Humes}
\affiliation{Northern Arizona University \\
Flagstaff, Arizona \\
86011, USA}

\author[0000-0003-3091-5757]{Cristina A. Thomas}
\affiliation{Northern Arizona University \\
Flagstaff, Arizona \\
86011, USA}

\author[0000-0001-9265-9475]{Joshua P. Emery}
\affiliation{Northern Arizona University \\
Flagstaff, Arizona \\
86011, USA}

\author[0000-0002-8296-6540]{Will M. Grundy}
\affiliation{Northern Arizona University \\
Flagstaff, Arizona \\
86011, USA}
\affiliation{Lowell Observatory \\
Flagstaff, Arizona \\
86001, USA}

\begin{abstract}

The recently launched Lucy mission aims to understand the dynamical history of the Solar System by examining the Jupiter Trojans, a population of primitive asteroids co-orbital with Jupiter. Using the G280 grism on the Hubble Space Telescope's Wide Field Camera 3 we obtained near ultraviolet spectra of four of the five Lucy mission targets--(617) Patroclus-Menoetius, (11351) Leucus, (3548) Eurybates, and (21900) Orus--to search for novel spectral features. We observe a local reflectance minimum at 0.4 $\mu$m accompanied by an increase in reflectance from 0.35-0.3 $\mu$m in the spectra of Patroclus and Orus. We use the principles of Rayleigh scattering and geometric optics to develop a Hapke optical model to investigate whether this feature can be explained by the presence of submicroscopic grains on Trojan surfaces. The near ultraviolet ``bump" feature can be explained by scattering due to fine-grained opaques (iron, amorphous carbon, or graphite) with grain sizes ranging from 20 - 80 nm. 

\end{abstract}


\section{Introduction} \label{sec:intro}

The Jupiter Trojan asteroids, located near Jupiter's L4 and L5 Lagrange points, orbit in 1:1 mean motion resonance with Jupiter. The particular gravitational relationship between Jupiter and its Trojans has been leveraged to provide constraints on the early formation and migration history of Jupiter and other gas giants (eg. \cite{Marzari_1998, Morbidelli_2005, Nesvorny_2013, pirani_2019}). Any explanation of the emplacement of the Trojan population must account for the physical and orbital properties of the Trojan population as a whole. Thus, understanding the characteristics of the Jupiter Trojan population is of particular interest, as these measurements can provide additional constraints on dynamical models of gas giant formation and migration. Furthermore, as primitive D-, P-, and C- type asteroids (\cite{bendjoya-2004, roig-2008, DemeoCarry_2014}), the Jupiter Trojan population is thought to have undergone very little geothermal processing (such as aqueous alteration) since their emplacement within the Trojan clouds (\cite{Lazzarin_1995}). In contrast, geological processing through collisions is thought to play an important role in the evolution of Trojans, with the cutoff between primordial objects and collisional fragments occurring for diameters less than $\sim 90$ km \cite{BINZEL1992}.  The presence of collisional families within the Trojans (including the Eurybates family) provide additional evidence of the importance of collision processes in shaping the Trojan populations \cite{vinogradova2015, rozehnal2016}. The higher light curve amplitudes of collisional fragments in the Trojan swarms compared to Main Belt asteroids of the same sizes reported by \cite{BINZEL1992} may be due to the Trojans being composed of lower-strength material that is less resistant to collisional fragmentation than Main Belt asteroids.
Due to the primordial nature of this population and the potential of further investigation of these objects to answer questions about the early Solar System, the Jupiter Trojans are the targets of the recently launched Lucy mission (\cite{Levison_2021}). 

Spectroscopy is one line of investigation that provides insight into the nature of the Trojan population. Spectroscopic studies can constrain the chemical composition of Trojan surfaces, which provide clues to their original source population and the changes this population have undergone since their emplacement. In the visible and near infrared (VNIR) region of the spectrum, Trojan asteroids display moderately red sloped, low albedo, featureless reflectance spectra as reported in \cite{Emery_2004}.  Compared  to  other  populations  with  red  sloped  VNIR spectra, the spectral slopes of Jupiter Trojans are relatively moderate, comparable to populations such as the Neptune Trojans, as well as the more neutral-toned Centaurs and scattered disk Kuiper Belt objects (\cite{Sheppard_2006}). Among the Jupiter Trojans themselves, there is evidence of two distinct subpopulations: the R (red) and LR (less red) Trojans, which are distinguished by their VNIR spectral slopes (\cite{Emery_2010, Wong_2014}). The generally reddish and featureless VNIR slopes of Jupiter Trojans have been interpreted in the literature as due to the presence of anhydrous silicates and a small fraction of complex  organics (\cite{Emery_2004}). In particular, \cite{Emery_2004} favor pyroxenes with low (0-40\%) iron content. At longer wavelengths, \cite{Brown_2016} reports a 3.1 $\mu$m feature among LR population asteroids which may be due to either fine-grained water frosts or the N-H stretch. In the mid-infrared, the Trojan asteroids (624) Hektor, (1172) Aneas, (911) Agamemnon, and (617) Patroclus, 10- and 20- $\mu$m emission features have been interpreted as being due to fine grained silicates (magnesium rich olivines and pyroxenes) suspended in either a fairy-castle like structure or a medium transparent in this region of the infrared (e.g. \cite{Emery_2006, Mueller_2010, Yang_2013}). 

Given the limited constraints these features place on the compositions of Trojans, attention has recently turned from the relatively featureless VNIR region of the spectrum to the ultraviolet (UV). Prior ground based spectrophotometry (\cite{Karlson_2009, Zellner_1985}) of the Trojans indicated a potential absorption feature in this region shortwards of 0.4 $\mu$m. In addition, \cite{Yang_2013} posited that fine grained silicates suspended in a salt matrix could produce the 10-micron silicate emission feature. Similarly, \cite{Wong_2016} invoked H$_2$S as a coloring agent to explain the VNIR slope bimodality in Trojans. Both of these hypotheses predict an absorption feature in the UV, from salts in the case of the Yang et al. hypothesis (\cite{clark_library_2007}) and H$_2$S in the Wong \& Brown hypothesis (\cite{Wong_2019}). Exploring the H$_2$S hypothesis, \cite{Wong_2019} used the Hubble Space Telescope (HST) to measure UV spectra of a group of six Trojan asteroids. While \cite{Wong_2019} did not see evidence of either the salt or H$_2$S absorption in their spectra, their work showed that the R and LR group Trojans also looked distinct from each other in the UV. LR Trojans showed a redder, steeper slope in the ultraviolet than the R Trojans, a reversal of the trend seen in the VNIR. The authors attributed this trend to a difference in either iron or organic content between the two subpopulations (\cite{Wong_2019}). 

\begin{deluxetable*}{lcccccc}[!ht]
\tablenum{1}
\tablecaption{Key properties of observed asteroids. \label{tab:asteroids}}
\tablewidth{0pt}
\tablehead{
 &  &  & \colhead{Effective Diameter} &  &  & \colhead{Inclination}
 \\
\colhead{Name} & \colhead{Taxonomic Class} & \colhead{L$_n$} & \colhead{(km)} & \colhead{Albedo}   & \colhead{Eccentricity} & \colhead{(deg)}
 }
\startdata
(617) Patroclus-Menoetius & P & L5 & 140 & 0.047 & 0.139 &	22.05 \\
(11351) Leucus & D & L4 & 34.2 & 0.079 &0.064 &	11.56 \\
(3548) Eurybates &  C & L4 & 63.9 & 0.052 & 0.089 &	8.06 \\
(21900) Orus & D & L4 & 50.8 & 0.075 & 0.037 &	8.47 \\
\enddata
\tablecomments{Patroclus-Menoetius is a near equal-size binary: the given diameter is the effective spherical diameter based on NEOWISE observations (\cite{Mainzer_NEOWISE}). In \cite{Levison_2021}, the diameters of Patroclus and Menoetius are given as 113 km and 104 km, respectively. Taxonomic classes, eccentricities, and inclinations are also from \cite{Levison_2021}.}
\end{deluxetable*}

\section{Methods}
\subsection{Observations \& Data Reduction}

We used the Hubble Space Telescope (HST) to measure ultraviolet and visible spectra (0.2 - 1.0 microns) of four of the five Lucy mission targets (Eurybates, Leucus, Orus, and the Patroclus-Menoetius binary) during Cycles 25 and 26 as part of programs \#15259 and \#15504, as summarized in Tables \ref{tab:asteroids} and  \ref{tab:obs}. Spectra were obtained using the G280 grism on the Wide Field Camera 3 (WFC3) in slitless spectroscopy mode (R$\sim70$), along with contemporaneous imagery taken with the F300X filter (\cite{WFC3}). The FOV of the UVIS camera is a 162'' by 162'' rhombus when projected on the sky (\cite{WFC3_handbook}). Because the spectrograph operates in slitless mode and has a large field of view, background sources are also imaged: their contributions to the image are quantified and removed during reduction.

Spectra were reduced using the aXe slitless spectroscopy reduction pipeline (\cite{Kummel_2009}), with a slight modification to account for the non-sidereal motion of the targets across the sky. The pipeline uses Source Extractor (\cite{Bertin_1996}) on F300X images to locate all objects, then uses header information about the field of view of subsequent grism images to calculate the pixel location of each object in the generated catalog under the assumption that targets detected in the F300X images do not move in RA and Dec, though their pixel locations may move due to the motion of the telescope. Since our targets move relative to the fixed RA-Dec coordinate system, for each subsequent image we must calculate and correct for the non-sidereal motion of the targets. To calculate this shift, we measured the pixel location of a fixed point in RA and Dec in each F300X and G280 image, then corrected by the difference between these measurements (in pixels) between the F300X and G280 image for the catalog entry corresponding to the asteroid. This method relies on the assumption that the telescope is tracking the target, i.e. that the pixel location of the asteroid does not vary significantly between each image, and can be verified by visual examination of the extracted spectral traces of the asteroids produced by the pipeline. A properly corrected catalog entry is easily identified by an extracted spectral trace that is linear and centered on its brightest pixels.

After each object is located, the reduction pipeline calculates where the spectral trace of each object falls on the chip. The pipeline then extracts the spectrum by projecting a virtual ``slit'' onto the image, its width and position based on the output of Source Extractor. We used a linear interpolation of the ten pixels on either side of the ``slit'' to calculate the background contribution, as we did not have master sky images for background subtraction.

\begin{deluxetable*}{lcccccc}[!ht]
\tablenum{2}
\tablecaption{Summary of HST observations. \label{tab:obs}}
\tablewidth{0pt}
\tablehead{
\colhead{}  & \colhead{Observation Midpoint} & \colhead{V} & \colhead{Phase Angle} & \colhead{F300x} & \colhead{G280} & \colhead{Program}
 \\
\colhead{Object}  & \colhead{(UTC)} & \colhead{(mag)} & (deg)& \colhead{exp. time (s)} & \colhead{exp. time(s)}
 & \colhead{\#}
 }
\startdata
(617) Patroclus-Menoetius & 2018-03-31 10:31:23 & 16.07 & 5.49 & 50 & 1924 & 15259\\
(11351) Leucus & 2018-08-18 13:01:07 & 18.04 & 3.47 & 100 & 4228 &  15259\\
(3548) Eurybates &  2019-08-11 17:18:25 & 16.99 & 6.47 & 10 & 1848 & 15504 \\
(21900) Orus & 2019-09-03 11:47:11 & 17.19 & 4.26 & 50 & 1928 & 15504 \\
\enddata
\tablecomments{Exposure time refers to the total number of seconds, combining all integrations with the same filter. Note that all observations were taken within 1 orbit with the exception of Leucus, which required 2 orbits due it its relative faintness. Magnitudes and phase angles are from JPL's Horizons database.}
\end{deluxetable*}

Following spectral extraction with aXe pipeline, the data were combined to produce a single spectrum for each object. First, in order to remove potential noise from 0th order background sources that happened to intersect with the spectral trace, outliers were rejected from individual spectra by removing data points that exceeded 5 sigma from the 20 point moving mean box average spectrum. To convert from flux to relative reflectance, we used a solar spectrum from \cite{Meftah_2018}, smoothed by a 10-point wide Gaussian filter. Then, each spectrum fitted to the solar reference by applying a small ($<$10 nm) wavelength shift and multiplying by an overall scale factor to produce relative reflectance spectra normalized to unity at 500 nm. The shifting step was performed to account for sub-pixel wavelength shifts to ensure proper alignment of narrow solar lines in the UV. This step is analogous to fitting subpixel shifts to align and correct for telluric absorption in ground based data. The final combined spectrum was produced by taking the mean of all measurements within 5 sigma of the median reflectance value to reject artifacts that affect individual frames.

In addition to taking UV spectra with the Hubble Space Telescope, we also obtained visible (0.35 - 1.0 microns) spectra of (617) Patroclus and (21900) Orus using the Lowell Discovery Telescope's DeVeny spectrograph (\cite{DeVeny}). Observations were taken using the DV1 grating (150 lines/mm, R$\sim$450) and 3.0'' slit. Observing circumstances are summarized in Table \ref{tab:obs-ldt}. Observations of each Trojan were accompanied by observations of standard stars bracketing the asteroid observation. Spectra were processed using the SPECTROSCOPYPIPELINE (SP) developed by \cite{Devogele_SP} for the DeVeny spectrograph, which performs bias and flat correction, cosmic ray subtraction, wavelength calibration, telluric and solar correction. The extracted spectra were then smoothed and binned using a 10-point wide box filter.

\begin{deluxetable*}{lcccccc}[!ht]
\tablenum{3}
\tablecaption{Summary of LDT observations. \label{tab:obs-ldt}}
\tablewidth{0pt}
\tablehead{
\colhead{}  & \colhead{Observation Start} & \colhead{V} & \colhead{Phase Angle}& \colhead{exp. time} &  & 
 \\
\colhead{Object}  & \colhead{(UTC)} & \colhead{(mag)} & \colhead{(deg)} & \colhead{ (s)} & Airmass 
 & \colhead{Standard(s)}
 }
\startdata
(617) Patroclus-Menoetius & 2019-06-25 04:33 &  16.51 & 9.96 & 1800 & 1.47 & Land. 105-56 (G5V) \\
(21900) Orus & 2021-11-13 04:34 & 17.01 & 10.71 & 1200 & 1.07 & Land. 93-101 (G5V)  \\
& & & & & & HD 9986 (G2V) \\
\enddata
\tablecomments{Exposure time refers to the total number of seconds, combining all integrations. For Patroclus-Menoetius, no local standard was used, as the solar analog was at the same airmass as the asteroid. For Orus, both a local and solar standard were used to calibrate the reflectance spectrum. Spectral types for standard stars are from \cite{marsset2020} and the SIMBAD database \cite{simbad} Magnitudes and phase angles are from JPL's Horizons database.}
\end{deluxetable*}

\subsection{Hapke Modeling}

In order to investigate possible spectral features in our UV spectra, we constructed a model based on the Hapke model (e.g. \cite{Hapke_1981, Hapke_1993}) to produce simulated geometric albedo spectra at zero phase angle and without opposition surge. Our unscaled relative reflectance spectra obtained by HST were first scaled by a constant factor to convert them into geometric albedoes using data available from the literature. For (617) Patroclus, our reflectance measurements were scaled to the geometric albedo spectrum used in \cite{Emery_2004} using a simple chi-squared ($\chi^2$) minimization routine to splice the visible range of our observations to the existing VNIR spectrum of Patroclus, resulting in an albedo of 0.0471 at 550 nm. For (21900) Orus and (3548) Eurybates, we took the visual albedoes (0.075 and 0.052, respectively) reported in \cite{Grav_2012} as the values of the geometric albedo at 550 nm and scaled our observations accordingly. Similarly, we scaled the spectrum of (11351) Leucus to the visible albedo (0.037) reported in \cite{buie2021} at 550 nm. Using zero-phase angle geometric albedos as the output of our Hapke models both simplified the calculations required to obtain a spectrum and allowed us to incorporate albedo into our compositional fits, a constraint that is particularly important to include for relatively featureless spectra such as those of the Trojans in the VNIR as pointed out in \cite{Emery_2004}. 

Our spectral model simulates two different scattering regimes by calculating the single scattering albedo $w$ from the real ($n$) and imaginary ($k$) indices of refraction and particle diameter $d$. Following the methods in \cite{Fayolle_2021}, we calculate the single scattering albedo in two different regimes:  the Rayleigh scattering regime (Equations \ref{ray0} through \ref{ray3}) and the geometric optics regime (Equations \ref{geo0} through \ref{geo5}). For particles that are small compared to the wavelength of light ($\frac{\pi d}{\lambda} < 1$), we are in the Rayleigh scattering regime and obtain the single scattering albedo $w_{Rayleigh}$ (Equations 5.9, 5.13-5.14 in Ch 5. of \cite{Hapke_1993}) using

\begin{equation} \label{ray0} 
   w_{Rayleigh} = \frac{Q_S}{Q_E}, 
\end{equation}
where
\begin{equation} \label{ray1} 
    Q_E = \frac{24nk}{\left(n^2 + k^2 \right)^2 + 4\left( n^2 - k^2\right) + 4} X
\end{equation}

\begin{equation} \label{ray2} 
    Q_S = \frac{8}{3} \frac{\left((n^2 + k^2)^2 + n^2 - k^2 - 2 \right)^2+36n^2k^2}{\left((n^2 + k^2)^2 + 4(n^2 + k^2)+ 4\right)^2} X^4
\end{equation}

\begin{equation} \label{ray3} 
    X = \frac{\pi d}{\lambda}.
\end{equation}
For the Rayleigh regime, we capped the value of the single scattering albedo at 1, as a single scattering albedo larger than 1 is unphysical. For particles that are large compared to the wavelength of light (eg. $\frac{\pi d}{\lambda} \geq 1$), we are in the geometric optics regime and use the following equations to obtain the single scattering albedo $w_{Geometric}$. For particles that are large compared to the wavelength of light, $Q_E = 1$ \cite{Hapke_1981}, and thus $w_{Geometric} = Q_S$ (Equations 2.69, 5.42, 6.39-6.41 of \cite{Hapke_1993}). To calculate the the single scattering albedo in the geometric optics regime, we use

\begin{equation} \label{geo0} 
   w_{Geometric} = \frac{Q_S}{Q_E} = S_e + (1 -S_e), \frac{(1 - S_i)}{1 - S_i\Theta} \Theta 
\end{equation}
where
\begin{equation} \label{geo1} 
   S_e = \frac{(n-1)^2 + k^2}{(n+1)^2 + k^2} + 0.05 
\end{equation}

\begin{equation} \label{geo2} 
   S_i = 1 = \frac{4}{n(n+1)^2} 
\end{equation}

\begin{equation} \label{geo3} 
   \Theta = \exp({- \alpha \langle D \rangle} )
\end{equation}

\begin{equation} \label{geo4} 
   \langle D \rangle = \frac{2}{3} \left(n^2 - \frac{1}{n}(n^2 - 1)^{3/2} \right) d
\end{equation}

\begin{equation} \label{geo5} 
   \alpha = \frac{4 \pi k}{\lambda}. 
\end{equation}
For a modeled surface with $i$ compositional components, each with grain diameter $d_i$, mass mixing ratio $m_i$, and density $\rho_i$, the average single scattering albedo can be computed from the single scattering albedos $w_i$ of the individual components via the following equation (Equation 17 in \cite{Hapke_1981}), regardless of scattering regime,

\begin{equation} \label{avg w} 
   w = \frac{\sum_{i} \frac{m_i}{\rho_i d_i}w_i}{\sum_{i}\frac{m_i}{\rho_i d_i}}. 
\end{equation}
The single scattering albedo $w$ can be combined with the multiple scattering function $H$, backscattering function $B$, and phase function $P$ to produce the bidirectional reflectance (Equation 16 from \cite{Hapke_1981}) using 

\begin{equation} \label{bdrf} 
   r(\mu, \mu_0, g) = \frac{w}{4 \pi} \frac{\mu_0}{\mu_0 + \mu} \left((1 + B(g))P(g) + H(\mu_0)H(\mu) -1 \right). 
\end{equation}
Here, $\mu, \mu_0$ and $g$ are geometric factors with $\mu = \cos e$ and $\mu_0 = \cos i$, where $e, i$ are the angles of exitance and incidence, respectively. To simplify our model, we took $B = 0$ and $P = 1$, as in the case for isotropic scatterers with no opposition effect. These assumptions were shown in \cite{Emery_2004} to result in effective models of Trojan surfaces in the VNIR. The $H$ (Equations 8, 10 from \cite{Hapke_1981}) function is approximated as 
\begin{equation} \label{Hfun} 
   H(\mu) = \frac{1 + 2\mu}{1 + 2 \gamma \mu},
\end{equation}
where
\begin{equation} \label{gamma} 
   \gamma = \sqrt{(1 - w)}.
\end{equation}

\begin{deluxetable*}{lcl}[ht!]
\tablenum{4}
\tablecaption{Hapke model compositional compnents and their densities \label{tab:comps}}
\tablewidth{0pt}
\tablehead{
\colhead{Composition} & \colhead{Density (g/cm$^3$)} & \colhead{Source(s)} \\
 }
\startdata
H$_2$O Ice & 0.82 & \cite{Warren_1984, Westley_1998} \\
CO$_2$ Ice & 0.98 & \cite{Warren_1986, Luna_2012}\\
NH$_3$ Ice & 0.47 & \cite{Martonchik_1984, Luna_2012}\\
Fe & 7.874 & \cite{Cahill_2012} \\
Pyroxene 2 ($f_{Mg} = 1$) & 3.21 & \cite{Dorschner_1995, Emery_2004}\\
Pyroxene 4 ($f_{Mg} = 0.8$) & 3.36 & \cite{Dorschner_1995, Emery_2004}\\
Pyroxene 5 ($f_{Mg} = 0.7$) & 3.43 & \cite{Dorschner_1995, Emery_2004}\\
Pyroxene 6 ($f_{Mg} = 0.6$) & 3.51 & \cite{Dorschner_1995, Emery_2004}\\
Olivine 1 ($f_{Mg} = 0.5$) & 3.81 & \cite{Dorschner_1995, Emery_2004}\\
Olivine 2 ($f_{Mg} = 0.4$) & 3.92 & \cite{Dorschner_1995, Emery_2004}\\
Titan Tholin & 1.5 & \cite{Khare_1984} \\
Tholin/H$_2$O Mix & 1.5 & \cite{Khare_1993}\\
Graphite & 2.26 & \cite{Draine_1985}\\
Amorphous Carbon & 1.85 & \cite{Rouleau_1991} \\
\enddata
\tablecomments{For entries with multiple citations, the first citation is the source of the optical constants followed by the source of the density. In cases with only one citation, the source contains both optical constants and material densities with the exception of the tholin mixtures, for which we assume a density of 1.5 following \cite{Emery_2004} and iron, for which the density of elemental iron was used. For the olivines and pyroxenes, we adopt the labeling scheme of \cite{Emery_2004} to keep track of silicates with different magnesium-iron ratios. For these silicates, the magnesium fraction $f_{Mg}$ gives the fractional magnesium content of each sample such that the chemical formula for the pyroxenes is given by Mg$_{f_{Mg}}$Fe$_{1-f_{Mg}}$SiO$_3$ and the olivines by Mg$_{2f_{Mg}}$Fe$_{2-2f_{Mg}}$SiO$_3$}
\end{deluxetable*}

To obtain the geometric albedo at a phase angle of zero from the bidirectional reflectance function, a numerical integration of the bidirectional reflectance function over 100 rings was performed by  subdividing a spherical surface into areas with $\mu_i = \mu_e$, then summing the contribution to reflectance from each ring, weighted by projected cross sectional area as viewed face on.

The sources of optical constants ($n$ and $k$) for the surface compositions used in the Hapke model are summarized in Table \ref{tab:comps}. Optical constants were interpolated to the wavelengths of our observations. For certain materials with $k$ values that span several orders of magnitude, values of the real part of the optical constants were interpolated linearly in $\log \lambda$, $n$ space, and the imaginary part of the optical constants were interpolated linearly in $\log \lambda$, $\log k$ space as noted in \cite{Warren_1984, Warren_1986}, otherwise, optical constants were linearly interpolated in $\lambda$, $n$ and $\lambda$, $k$ space.  

The compositions used to model the Trojan spectra were drawn from the results of previous Hapke modeling of Trojan spectra in the VNIR (eg. \cite{Emery_2004}). In addition to the silicate, organic, and tholin compositions suggested by previous literature, additional ice species (H$_2$O, CO$2$, and NH$3$) were chosen as potential Rayleigh scatterers based on laboratory measurements by \cite{Hapke_1981_frost}. Ices were also chosen for their potential to explain the 3.1 micron absorption feature. \cite{Brown_2016} suggested the feature could be due to either very fine grained water frosts or species with the N-H stretch. Thermal modeling of the Trojans by \cite{GUILBERTLEPOUTRE2014} has indicated that water ice may be present as a surface or near-surface component at depths ranging from $\sim 10$ cm to $\sim$ 10 m depending on albedo, thermal inertia, and orbital characteristics of the asteroid. For low-obliquity Trojans, it is also possible that water ice is present at the surface near the poles. Thus, water ice is a plausible surface component to include in our modeling efforts. We include CO$_2$ and NH$_3$ as possible compositions due to the spectral similarities of Jupiter Trojans to comets (\cite{YE2016, KELLEY2017}), though we note that in pure ice form CO$_2$ and NH$_3$ are likely not currently stable on or near Trojan surfaces in the present day (\cite{Wong_2016}). Though altered residues from these ices may be present as surface components of Trojans (\cite{Brown_2016}), our model is limited in that it can only test for the presence of the unaltered ices.
Additional opaque species, including iron and carbon, were also chosen as Rayleigh scatterers due to their hypothesized roles in space weathering and particular features in the UV (see  \cite{Pieters_2000, Cloutis_2008, Hendrix_2016}.) 

To obtain best-fits, the model was fed an initial ``guess" composition based on adding a Rayleigh scattering species to the compositions of the best fit models in \cite{Emery_2004}. The initial guess was then refined by allowing the particle diameters and mass mixing ratios (mass fraction relative to the abundance of the first composition, which was fixed at a mixing ratio of 1) to vary until the reduced $\chi^2$ value of the difference between data and model was minimized. The model can also evaluate the $\chi^2$ metric for any input composition without attempting to determine the best fit in order to quantify how changing variables such as particle size affects the shape of the resulting spectrum. For this modeling effort, we focused on three-component surface models, generally consisting of silicates, a Rayleigh scatterer, and an additional component. The addition of a fourth component did not significantly improve the goodness of the model fit.

\section{Results}

\begin{figure}[!ht]
\plotone{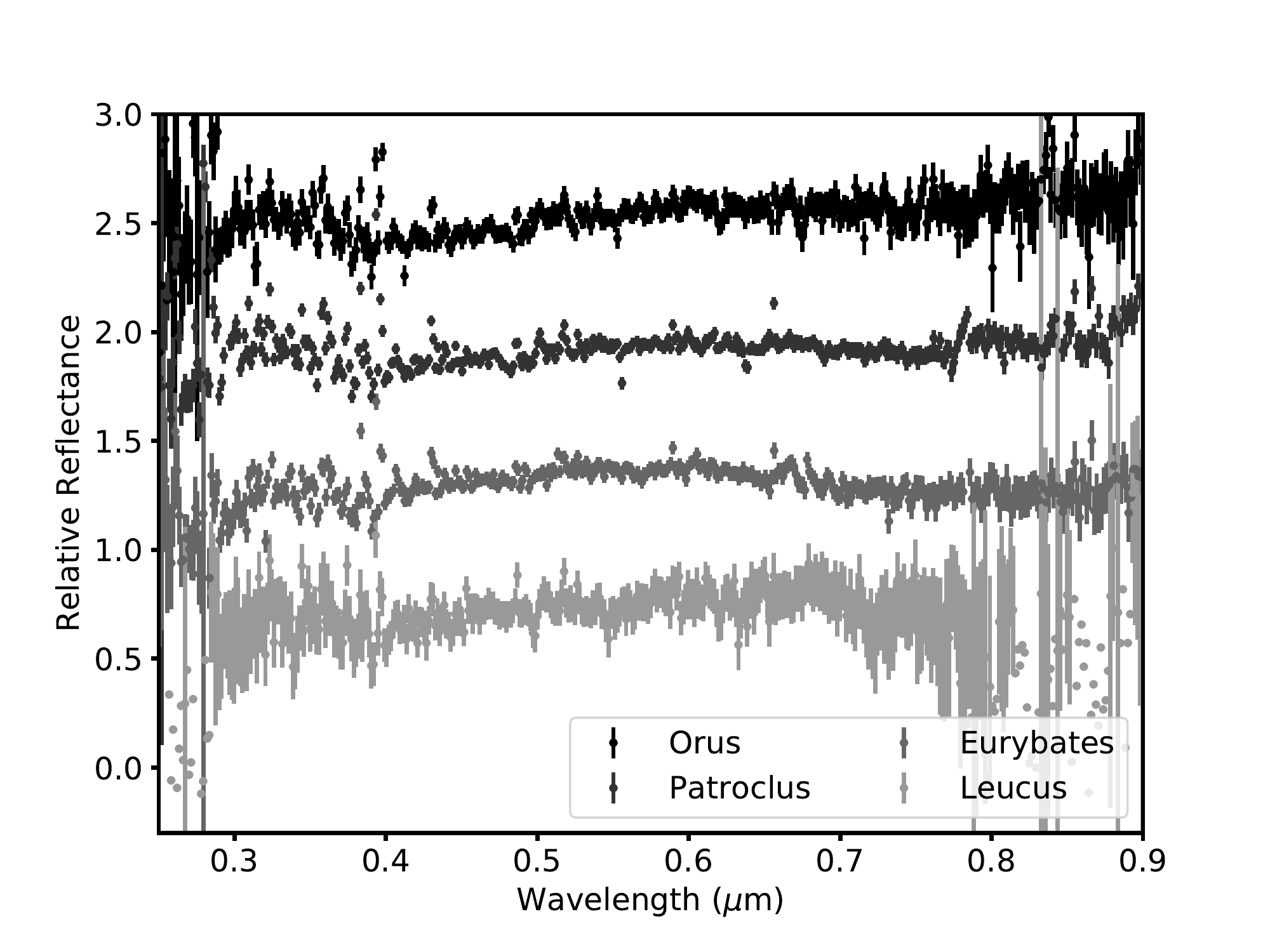}
\caption{Ultraviolet and visible spectra of Orus, the Patroclus-Menoetius binary, Eurybates, and Leucus from the Hubble Space Telescope. Spectra are normalized to 0.5 $\mu$m and offset for clarity.  \label{fig:four-UV}}
\end{figure}

\begin{figure}[!ht]

\gridline{\fig{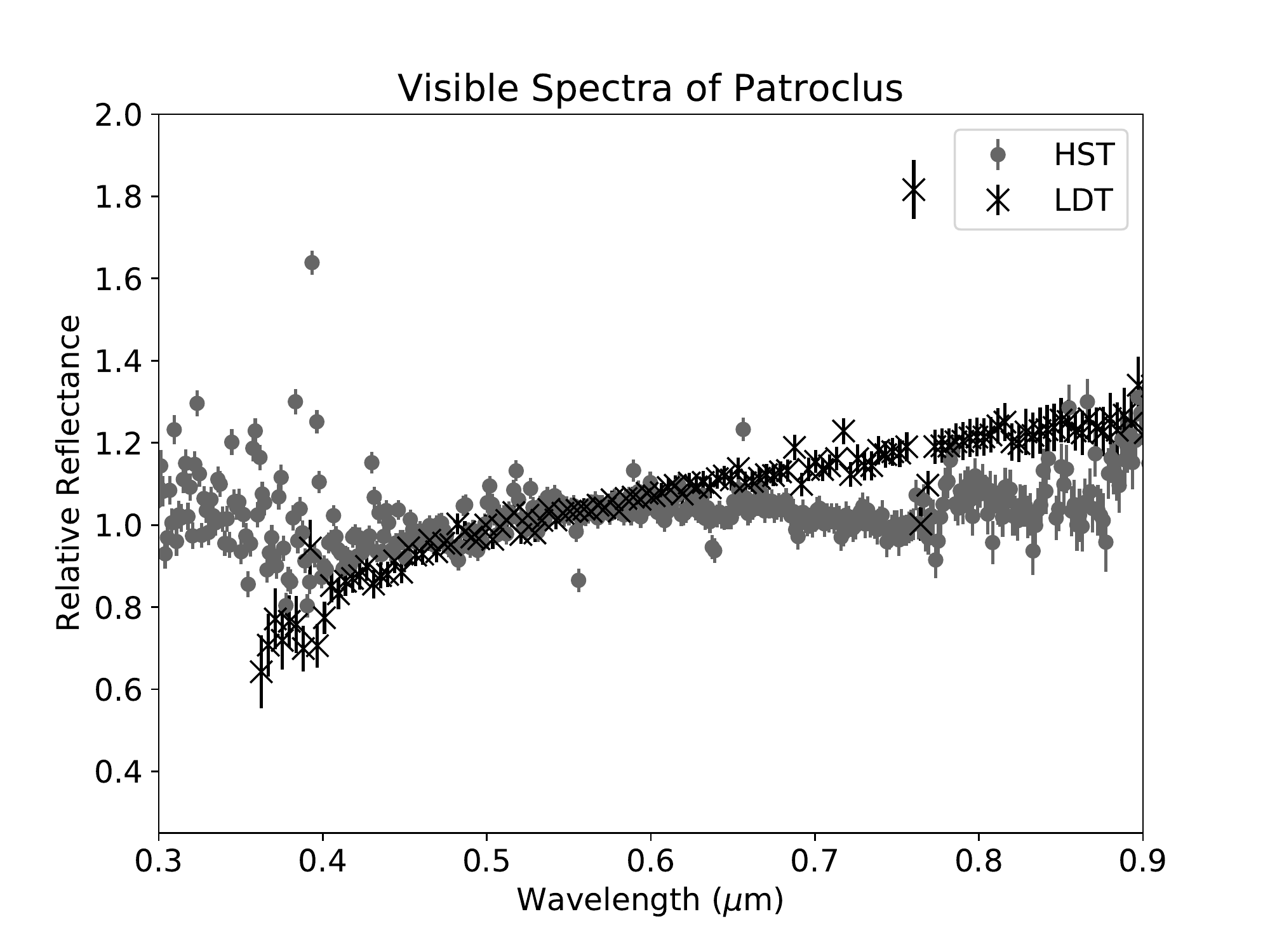}{0.5\textwidth}{(a)}
          \fig{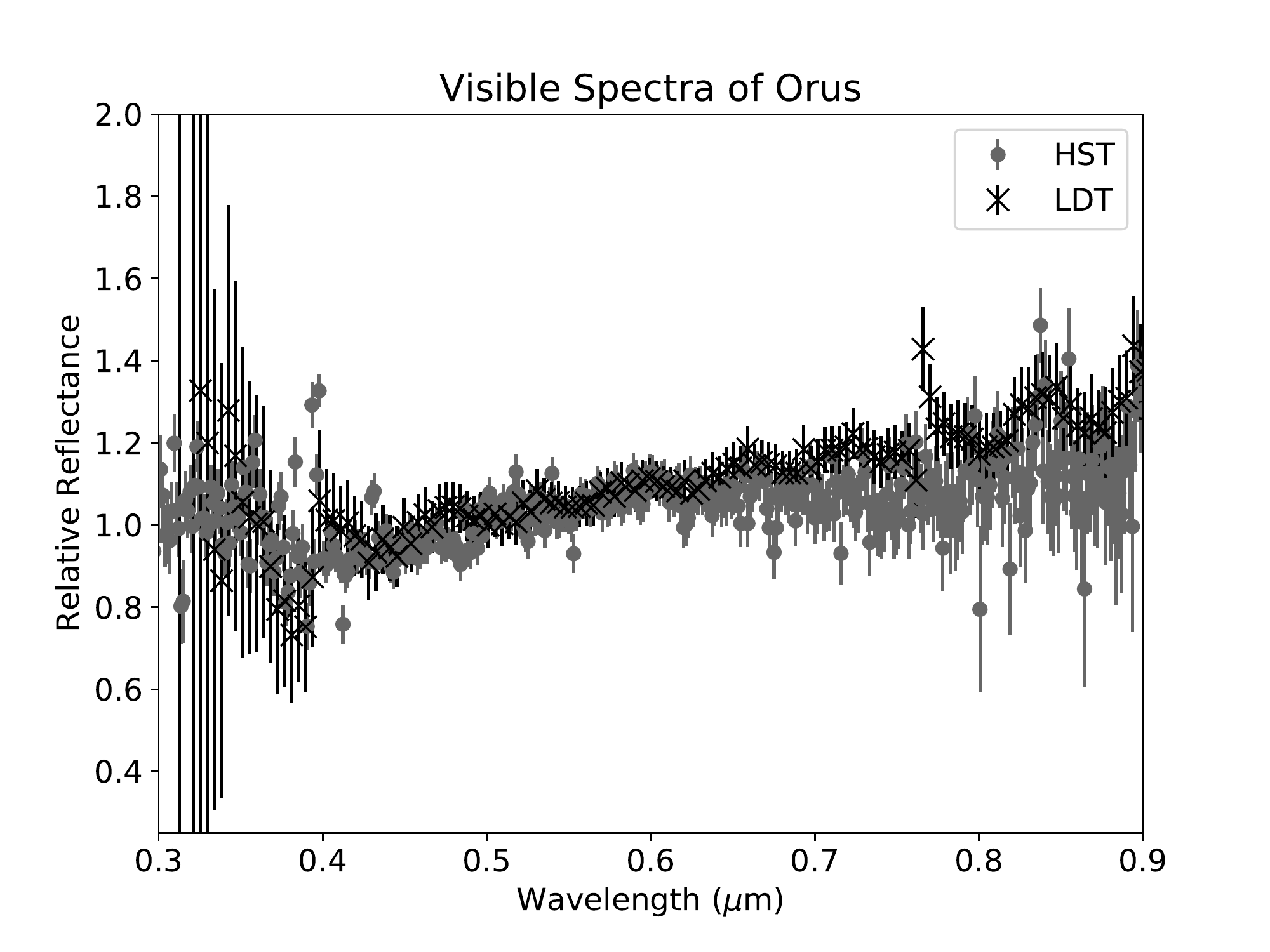}{0.5\textwidth}{(b)}
          }
\caption{Visible spectra of Patroclus (a) and Orus (b) as observed by the Hubble Space Telescope (HST) and the Lowell Discovery Telescope (LDT). Spectra were normalized at 0.5 $\mu$m. The observed slopes of Patroclus in the 0.4-0.55 $\mu$m region agree for the two observations, but the HST spectrum shows a flattening in reflectance starting at ~0.55 $\mu$m that is not seen in the LDT spectrum. Because this feature has not been reported in the spectrum of Patroclus before, it may be due to instrumental error or rotational variability (see text). In the same wavelength region, the two spectra of Orus are in better agreement. \label{fig:two-pats}}
\end{figure}

The spectra of our four targets, normalized to 0.5 $\mu$m and offset for clarity are shown in Figure \ref{fig:four-UV}. While the visible spectra of the Trojans as observed by Hubble are fairly consistent with previous observations of these targets, we note that in the spectrum of Patroclus and Eurybates, there is a shallowing of slope and subsequent downturn in reflectance from 0.6-0.7 $\mu$m, as compared to the steeper slopes of the 0.4-0.6 $\mu$m region of the spectrum.
Similar features have been reported in the spectrum of Eurybates (\cite{SouzaFeliciano_2020}) and some Eurybates family members (\cite{Fornasier_2007}). However, visible spectra of Patroclus including those in \cite{Emery_2004} and our LDT spectra (See Fig \ref{fig:two-pats}) do not show this downturn. Given that Patroclus is not a member of the Eurybates family and is typically classified as a P type rather than a C type asteroid, we do not expect similar features to appear in the spectra of both Patroclus and Eurybates. This feature may be due to instrumental issues at the long wavelength end of the G280 grism range, as higher spectral orders begin to overlap with the brightest spectral trace, as noted in \cite{wakeford_2020}, though this explanation does not explain why this feature would be present in the spectra of some of our targets and not others. It is possible that this feature appears in the spectrum of Patroclus due to rotational variability in appearance, as seen with Eurybates by \cite{SouzaFeliciano_2020}.  To assess what portions of the surfaces of Patroclus and Menoetius are observed at different times, we define a lat,lon coordinate system for Patroclus based on the Patroclus-Menoetius circular mutual orbit solution from \cite{grundy2018upcoming}. The positive pole is defined by the angular momentum vector of the mutual orbit. Zero longitude is defined by the direction from Patroclus to Menoetius, which is constant if the mutual orbit has been tidally circularized, the two bodies' spins are tidally locked to their orbit, and no libration occurs.  The coordinate system assumes a spherical shape for Patroclus and is not distorted to account for its elongation into a triaxial ellipsoid. This coordinate system is analogous to that used for Pluto, where the mean direction to Charon defines zero longitude.  Accounting for light time, we then compute the sub-Earth latitude and longitude on Patroclus at the mean time of each observation as a way of evaluating what hemispheres of the bodies were oriented toward Earth at the time of the observations. Using this method, we find the sub-earth latitude and longitude for the HST observations are +10.9, 26.5 and the sub-earth latitude and longitude for the LDT observations are -8.1, 280.2. These observations are separated by $\sim 106^{\circ}$, lending support to the interpretation that the difference in Patroclus's long wavelength spectrum may be due to a difference in viewing geometry between the two observations. In both these geometries, both Patroclus and Menoetius are visible (i.e. neither is eclipsing the other), though we cannot resolve the individual components in our imagery.

\begin{deluxetable*}{lccc}[ht!]
\tablenum{5}
\tablecaption{Results of Monte Carlo test for the  ``bump'' feature \label{tab:monte-carlo}}
\tablewidth{0pt}
\tablehead{
\colhead{} & \colhead{Mean Slope} & \colhead{Slope Standard Deviation} \\
\colhead{Asteroid} & \colhead{($10^{-5}$ per $\mu$m)} & \colhead{($10^{-5}$ per $\mu$m)} & Bump?\\
 }                   
\startdata                          
(617) Patroclus-Menoetius & -6.3  & 1.3 & Yes \\
(11351) Leucus & 1.0 & 5.6 & No \\
(3548) Eurybates & 4.7 & 1.8 & No \\
(21900) Orus & -6.0 & 2.8 & Yes \\
\enddata
\tablecomments{For each asteroid, 10,000 synthetic spectra were produced and slopes fit to the 0.3-0.4 $\mu$m region to estimate the slope and slope error in the near UV (see text). We consider a spectrum to show the ``bump'' feature if an asteroid has a negative slope significant at the 2-sigma level.}
\end{deluxetable*}

In the ultraviolet, the spectra do not show evidence of absorption features, including the salt feature that would support the hypothesis of \cite{Yang_2013} or the H$_2$S feature predicted by \cite{Wong_2016}. However, the spectra of Orus and Patroclus show a local reflectance minimum at around 0.4 $\mu$m, followed by an increase in reflectance from 0.3 - 0.35 $\mu$m, which we refer to as the ``bump" feature. To determine whether an asteroid spectrum showed evidence of the bump feature, we used a Monte Carlo approach. For each asteroid, we produced 10,000 synthetic spectra by drawing reflectance values for each wavelength from a Gaussian distribution with a mean equal to the reflectance and standard deviation equal to the error. Then, for each synthetic spectrum, we fit a line to the 0.3-0.4 $\mu$m region and measured its slope. Applying this method to all four asteroids (See Table \ref{tab:monte-carlo}), we find that Orus and Patroclus have negatively (e.g. blue) sloped reflectances at a confidence level of 2 sigma. Using this same method, Eurybates has a positively sloped (e.g. red) reflectance spectrum in this region at the 2 sigma level, while Leucus has an ambiguous or neutral slope at the 2 sigma level.

\begin{figure}[!ht]
\plotone{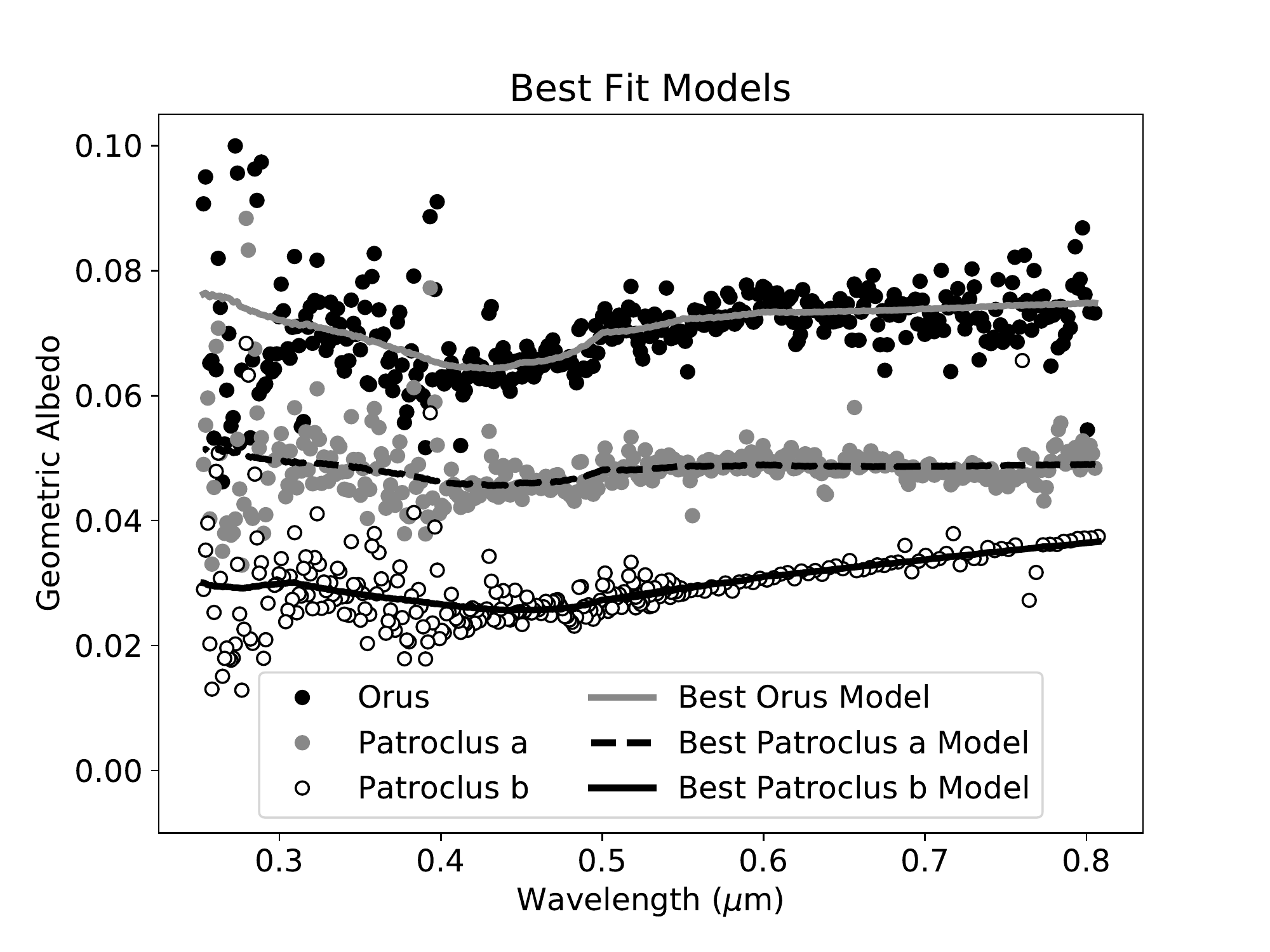}
\caption{Best fit models for Orus and Patroclus (solid and dashed lines) plotted alongside geometric albedo data from HST and LDT observations (filled and open circles). The spectrum for Patroclus a uses only HST data, while the spectrum for Patroclus b uses HST data for short wavelengths ($<0.55 \mu $m) spliced with LDT data at longer wavelengths ($>0.55 \mu$ m). Patroclus b is offset for clarity. See Table \ref{tab:bestfits} for details on the best fit models.  \label{fig:best-fits}}
\end{figure}

We used Hapke modeling in order to investigate whether this potential new feature in the near-UV can be explained by Rayleigh scattering due to fine grained particles on the surfaces of these two asteroids. Because Orus and Patroclus show this feature according to our Monte Carlo analysis, we focus our analysis on these two asteroids in particular. To account for the difference in Patroclus spectra from the HST and LDT, we fit models to two different versions of the spectrum: the unaltered spectrum from the Hubble Space Telescope (Patroclus a); and a spectrum constructed by splicing together short wavelength HST spectra and long wavelength LDT spectra at 0.55 $\mu m$ (Patroclus b). We fit our model to the 0.25 - 0.8 micron range, excluding regions with large amounts of noise and point-to-point scatter. Our five best fitting models, as evaluated by minimum reduced $\chi^2$, are summarized in Table \ref{tab:bestfits}. In general, Orus is well modeled by a three component spectral model consisting of a pyroxene, an opaque Rayleigh scatterer, and a third darkening component, the composition of which can vary. Patroclus shows consistently worse-fitting models, which may be due to spurious features in its visible spectrum, or its relatively smaller error bars. 

\begin{figure}[!ht]
\gridline{\fig{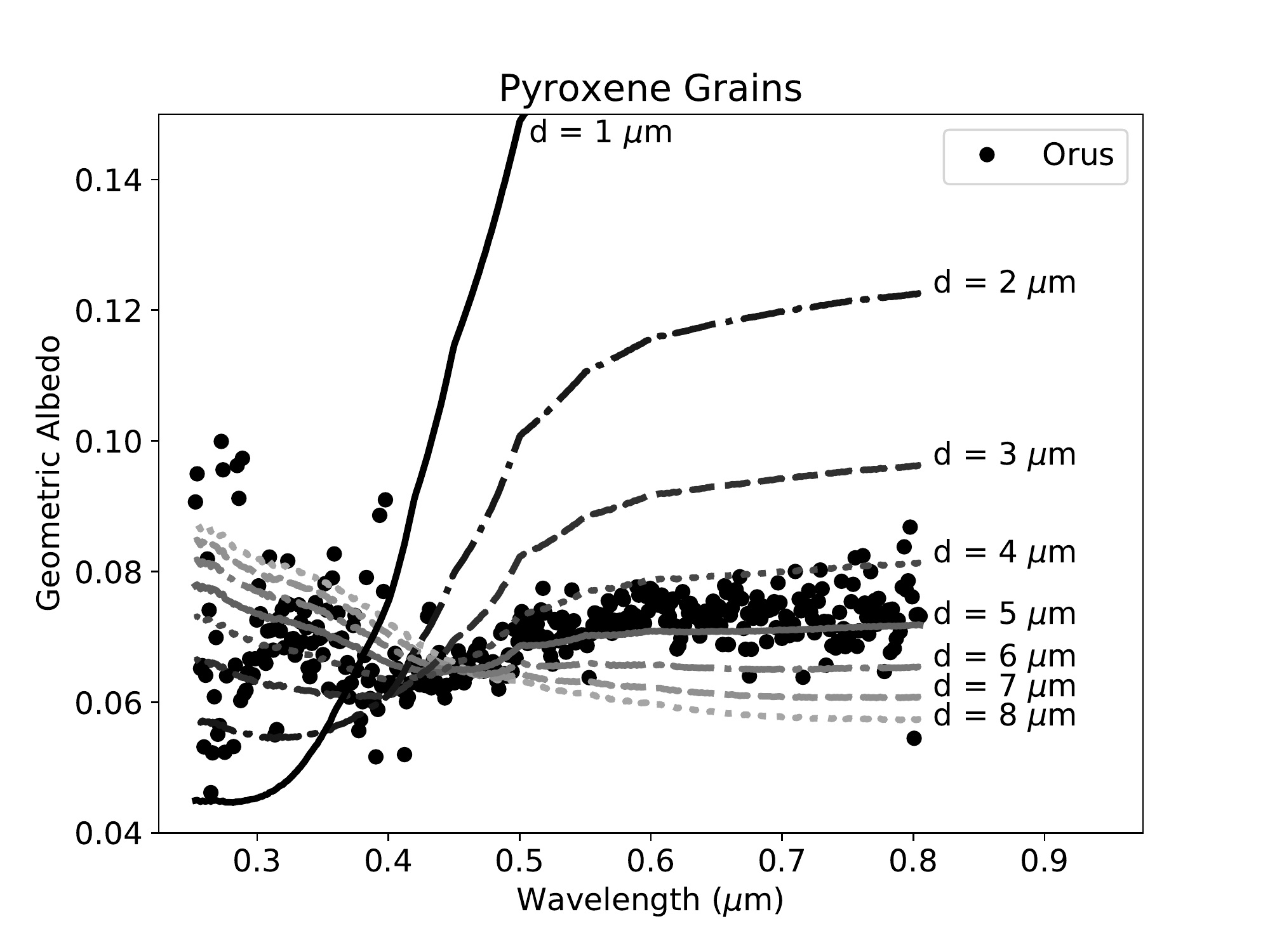}{0.5\textwidth}{(a)}
          \fig{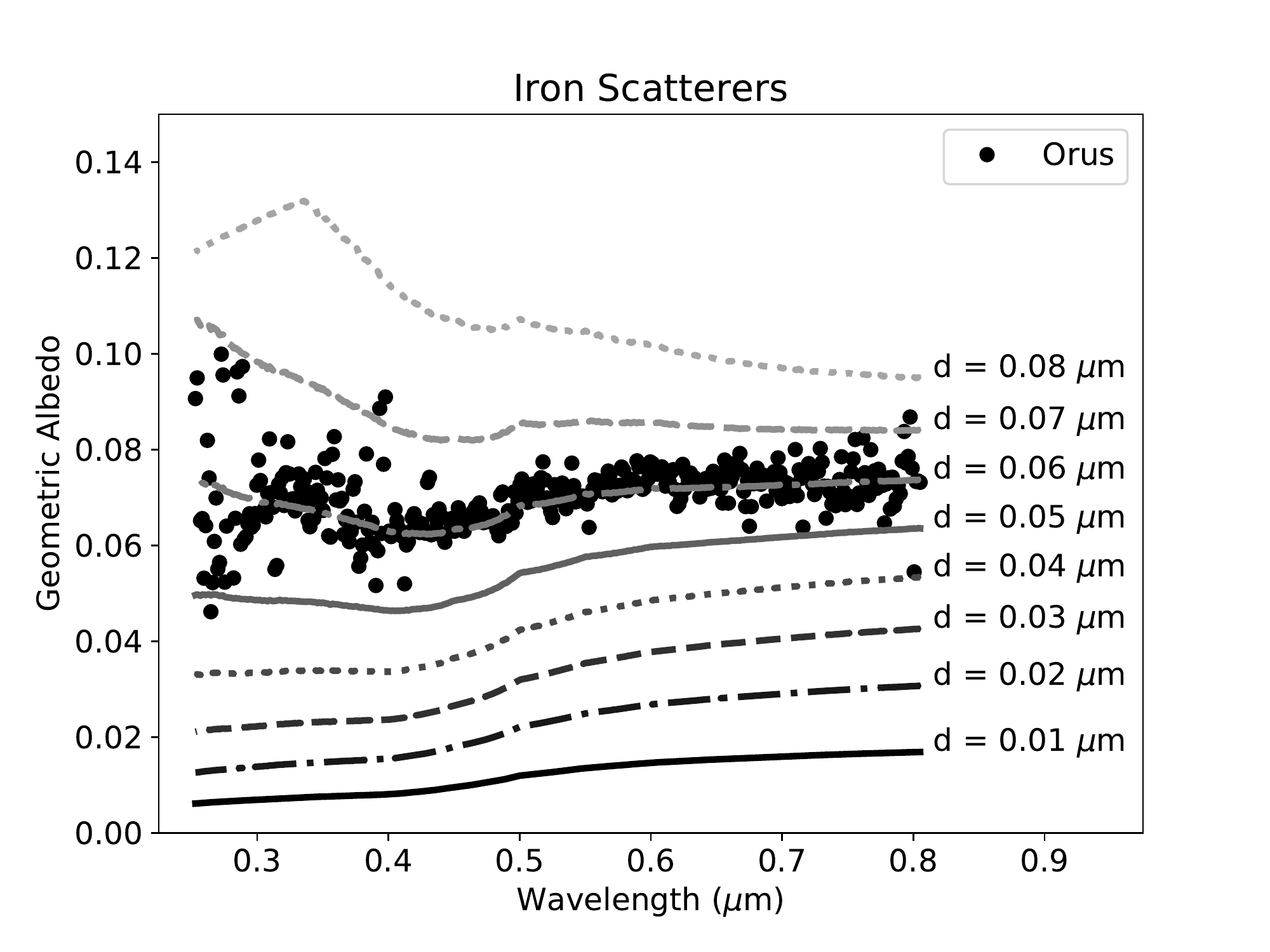}{0.5\textwidth}{(b)}
          }
\caption{The effect of changing grain sizes of different components of the best fit model for Orus (See Table \ref{tab:bestfits}, Orus 1). Changing the grain size of the pyroxene in this model (panel a) controls the magnitude of the slope in the visible and the extent of absorption in the UV. Changing the grain size of the iron scatterers in this model (panel b) controls the spectral contrast of ``bump" feature in the UV, as well as the wavelength at which the spectrum begins to brighten. The spectrum of Orus is also shown for reference purposes, without error bars for clarity--note that the scale used to display the finer details of the spectra cuts off one prominent outlier. \label{fig:grain-sizes}}
\end{figure}

For all of our best fit models (e.g. Fig \ref{fig:best-fits}), Pyroxene 5 and Pyroxene 2 are the most prominent silicates in the spectrum. The grain size of the pyroxenes is largely responsible for setting the gradient of the spectral slope in the visible: larger grain sizes result in stronger absorption in the blue, and steeper slopes. This finding is consistent with the results of \cite{Emery_2004}, which found that silicates mixed with low albedo materials (such as organics) can reproduce the red slopes and low geometric albedos of the Trojans. The grain size of the Rayleigh scattering species is responsible for the spectral contrast and wavelength at which the ``bump" feature begins to appear (see Figure \ref{fig:grain-sizes}). Similarly, the grain size and abundance of the third component in our models helps to set the overall geometric albedo. This component is generally a darkening agent like iron or amorphous carbon, or an additional pyroxene. In general, while Pyroxene 5 was present in all the best fit spectra presented here, silicates of other compositions, including other pyroxenes and olivines, could be substituted as the primary silicate component, though these compositions did not match the observed spectra as well. The fine grained Rayleigh scatterers in our models are all opaques (eg. iron, carbon), rather than ices. While our best fitting models use iron as a Rayleigh scatterer, amorphous carbon and graphite can both be substituted as Rayleigh scatterers to provide a good fit. Though not the main focus of our modeling efforts, we also find that the inclusion of 0.01-0.02 (mass mixing ratio, compared to pyroxene) of iron Rayleigh scatterers can improve compositional models of Leucus in particular, but the inclusion of Rayleigh scatters is not required to explain the spectral shapes of Leucus and Eurybates among the best fit models for these asteroids. 

\movetabledown=2.6in
\begin{rotatetable}

\begin{deluxetable*}{llccc||llccc||llccc}
\tablenum{6}
\tablecaption{Best fit optical models and their compositions  \label{tab:bestfits}}
\tablewidth{0pt}
\tablehead{
\colhead{Model} &\colhead{Comp.} & \colhead{Particle} & \colhead{Mixing} & \colhead{Reduced } & \colhead{Model} &\colhead{Comp.} & \colhead{Particle} & \colhead{Mixing} & \colhead{Reduced } & \colhead{Model} &\colhead{Comp.} & \colhead{Particle} & \colhead{Mixing} & \colhead{Reduced }  \\
& & \colhead{Diam. ($\mu$m)} & \colhead{Ratio}
  & \colhead{$\chi^2$} & & & \colhead{Diam. ($\mu$m)} & \colhead{Ratio}
  & \colhead{$\chi^2$} &&  & \colhead{Diam. ($\mu$m)} & \colhead{Ratio}
  & \colhead{$\chi^2$}\\
  }
\startdata
Orus 1 &	Pyroxene 5 &	4.64 &	1.00 &	1.67 & Pat. 1a & Pyroxene 5 &	4.38 &	1.00 & 6.37 &
Pat. 1b &	Pyroxene 5 &	6.60 &	1.00 &	8.05\\
&	Iron &	0.06 &	0.05	& & & Iron &	0.05 &	0.09 & & & Iron &	0.08 & 0.02\\
&	Iron &	0.43 &	0.14 & & & Iron & 1.84 & 0.93 & & & Amorph. C&	0.29 &	0.12 \\
\hline
Orus 2	& Pyroxene 5 & 4.66	 & 1.00 &	1.68 & Pat. 2a &	Pyroxene 5 &	3.53 &	1.00 & 6.39 & Pat. 2b &	Pyroxene 5 &	10.5 &	1.00 &	8.08\\
&	Iron &	0.06 &	0.05	& && Iron &	0.08 &	0.04 & & & Amorph. C &	0.04 &	0.01\\
&	Pyroxene 2 &	13.1 &	1.34 & & & Amorph. C	& 0.92 &	0.62 & & & Iron &	0.60 &	0.22\\	
\hline
Orus 3 &	Pyroxene 5 &	4.19 &	1.00 &	1.70 & Pat. 3a&	Pyroxene 5 &	4.33 &	1.00 & 6.43 & Pat. 3b	&Pyroxene 6 &	0.80 &	1.00 & 8.09\\
&	Iron &	0.08 &	0.03	& && Iron &	0.04 &	0.12 & & & Iron & 0.08 &	0.19 \\
&	Amorph. C &	4.51 &	0.54 & && Pyroxene 2 &	0.84 & 	0.25 & & & Amorph. C &	0.43 &	1.38 \\
\hline
Orus 4 &	Pyroxene 5 &	4.02 &	1.00 &	1.73 & Pat. 4a &	Pyroxene 5 &	3.98 &	1.00 & 6.53 &  Pat. 4b	& Pyroxene 5 &	10.6 &	1.00 &	8.11 \\
&	Amorph. C &	0.04 &	0.02	& & & Amorph. C & 0.03 & 0.06 & & & Graphite &	0.05 &	0.01\\
&	Pyroxene 2	& 10.2 & 	4.90 & & & Pyroxene 2 &	0.25 &	0.29 & & & Iron &	0.32 &	0.11 & \\	
\hline
Orus 5 &	Pyroxene 5 &	4.02 &	1.00 & 	1.73 & Pat. 5a &	Pyroxene 5 &	3.63 &	1.00 & 6.56 & Pat. 5b	& Pyroxene 6 &	1.19 &	1.00 &	8.13 \\
&	Graphite &	0.06 &	0.02	& & & Graphite & 	0.04 &	0.06 & & & Amorph. C &	0.04 &	0.08 \\
&	Iron &	1.86 & 	2.50 & && Iron &	0.86 &	2.06 &	& & Iron &	0.65 &	3.13 \\
\enddata
\tablecomments{The top 5 best fitting models for Orus and Patroclus. We present two versions of the Patroclus spectra: Pat. a models were fit to Hubble Space Telescope data; while Pat. b models were fit to the Hubble Space Telescope data at short wavelengths ($<0.55 \mu m$) spliced with data from the Lowell Discovery Telescope at long wavelengths ($>0.55 \mu m$). For each model we give the compositional components, their diameters, and mixing ratios. Amorphous carbon is abbreviated in this table as Amorph. C.  }
\end{deluxetable*}

\end{rotatetable}

To quantify error in particle diameter and mixing ratios, we generated a large number of synthetic spectra drawn from Gaussian distributions for each data point with a mean equal to the observed geometric albedo and standard deviation equal to the error in albedo. Then, the best fit model was fitted to each synthetic spectrum using the same model components to determine the scatter in particle diameter and mixing ratio. The resulting differences in particle diameter and mixing ratio only appeared in the third or fourth significant figures of each model parameter. This error is negligible compared to systematic errors arising from model assumptions. 

In order to better understand the range of possible model parameter values, we vary model parameters individually and compute how the spectrum and value of $\chi^2$ changes when the parameter is varied by small ($\sim$1-10\%) increments about the best fit model. In particular, we are interested in the range of possible sizes for the Rayleigh scattering component, which is responsible for the shape of the spectrum in the UV, as other modeling efforts have addressed the range of possible compositions responsible for the visible and near infrared spectra of Jupiter Trojans (\cite{cruikshank2001, Emery_2004, sharkey_2019, gartrelle_2021}). We focus on the five best-fit models of Orus in order to understand the range of possible Rayleigh scatterer sizes due to Orus's overall better fit to the model. We identified the Rayleigh scatterer grain sizes for which each model has a $\chi^2$ statistic of 2 or smaller, since all models with a $\chi^2$ statistic less than 2 produce optical models with geometric albedos that fall within the range of the scatter observed in the spectrum of Orus. Using this method, we find that the Rayleigh scatterers can be 0.001 - 0.002 $\mu$m larger or smaller and still provide a reasonable fit.

We note that the main source of error in our model is due to model limitations. In reality, Trojan surfaces are much more complex than a simple three component model. There will be grain size variation among each component and surface in-homogeneity, which is not accounted for in our model.

\section{Discussion}

In the most general terms, the compositional mix of our best fit models include a silicate, darkening agent, and a Rayleigh scatterer. This mixture agrees with previous modeling efforts for Jupiter Trojans and dark (D type) asteroids. Both \cite{Emery_2004} and \cite{gartrelle_2021} find that Trojans are well modelled by silicates mixed with opaques. However, unlike \cite{Emery_2004} and \cite{gartrelle_2021}, we include iron on its own as a darkening agent. \cite{cruikshank2001} report including iron sulfide (as an analog for metallic iron) as a component in their modeling but find no reason to prefer FeS as a component in comparison to their magnesium-rich pyroxene models.
Like \cite{Emery_2004}, we find that the red slopes of the Trojans can be accounted for with silicates alone, without needing to invoke organic components (e.g. tholins) in large quantities, in contrast with \cite{sharkey_2019}, which used Triton tholins as a primary component. In \cite{gartrelle_2021} phylosilicates provide the best fits for D type asteroid spectra, with opaque species masking characteristic absorption features. We were unable to test phylosilicates as silicate materials since optical constants in the near-UV are not publicly available for relevant phylosilicate species. However, like \cite{Emery_2004}, we find that while other silicate species can be substituted for pyroxene and still produce a reasonable fit, the best fits contained anhydrous silicates. Whether the Trojans are composed of hydrous or anhydrous silicates is an open question which this study does not have sufficient scope to answer, but may be addressed by the recently launched Lucy mission (\cite{Levison_2021}). Silicates have previously been observed in the spectra of Trojans; in particular, the silicate emission features at 10 and 20 microns have been noted by \cite{Emery_2006} and \cite{Mueller_2010}.

Our spectra also show evidence of a new spectral feature in the near-UV--the ``bump'' that is present in the spectra of Orus and Patroclus. Based on the spectra presented here, the presence of this ``bump" feature is not correlated with VNIR spectral properties of these asteroids: Orus is a member of the R group and Patroclus is a member of the LR group and both display this feature. Leucus is an R group asteroid and Eurybates is a more neutral-toned C type asteroid and neither asteroid displays the feature.
Our findings contrast with the findings of \cite{Wong_2019}, who examined a sample of six Trojan asteroids ((1143) Odysseus, (624) Hektor, (3451) Mentor, (659) Nestor, (911) Agamemnon, and (1437) Diomedes). \cite{Wong_2019} observed that slope features in the UV show a correlation to VNIR slopes: redder asteroids in the VNIR tended to have shallower slopes in the UV than their less-red counterparts, which showed a steepening in spectral slope starting around 0.45 $\mu$m. Not only do we not observe this slope reversal trend among our sample, we do not observe the ``bump'' feature in the spectra of \cite{Wong_2019}, though their spectra examine the same region at sufficient resolution to detect the ``bump.'' Applying the same Monte Carlo approach to quantifying the ``bump'' feature to the spectra from \cite{Wong_2014}, we find all asteroids in their sample are positively sloped (e.g. red) in the 0.4-0.3 $\mu$m region of the spectrum at the 2 sigma level. Taken together, our results and those of \cite{Wong_2019} suggest a diversity of UV spectral features among the Trojans.

Based on the results of Hapke modeling, we hypothesize that the ``bump" feature we observe in the spectra of Orus and Patroclus is due to Rayleigh scattering from fine grained (20-80 nm), opaque particles consistent with iron or carbonaceous compositions. Rayleigh scattering has been invoked to explain the spectral properties of planetary surfaces.
The conditions necessary for observing the bluing effect of small particles on the spectra of planetary surfaces was explored in \cite{BROWN_2014}, which found that Rayleigh scattering induced bluing can overtake the effects of absorption for sufficiently small, dark particles. In \cite{clark_2008} and \cite{clark_2012}, a Rayleigh scattering peak coupled with an ultraviolet absorber was shown to explain the spectral properties of dark materials on the Saturnian moons Dione and Iapetus. In those papers, the dark material is presumed to be exogenous in origin, and is present either embedded within or as a thin layer on top of the icy surfaces of the satellites. Only a small weight percentage is needed for Rayleigh scattering to affect the spectrum. 

However, unlike in our model, the Rayleigh scatterers in the Iapetus and Dione papers are embedded within an ice matrix. In our model, our scattering grains are mixed assuming separate particles averaged in an areal mixture, rather than as small grains embedded in a matrix. We implemented the option to test inclusions of fine grained materials within a silicate matrix using Maxwell-Garnett theory following methods outlined in \cite{Fayolle_2021}, but found that these models did not provide a good fit to the UV spectral features of interest. One key feature contributing to the shape of the Rayleigh peak in the spectra from \cite{clark_2008, clark_2012} is the transition between a regime dominated by Rayleigh scattering and a regime dominated by absorption of the iron particles. Due to large amounts of noise at the shortest wavelength end of our spectra, it is difficult for us to say if or where this transition happens for our Trojans, but given the results of \cite{Wong_2019} that showed the reddish slopes of the Trojans continue into the UV, we postulate that eventually the Rayleigh feature on the the Trojans could turn over as well, returning to the reddish slope trend at shorter wavelengths in the UV. 

In terms of size, the Rayleigh scattering species in our best fit models are referred to in the planetary literature as submicroscopic particles. Submicroscopic particles have been implicated in space weathering on airless bodies including Earth's moon (\cite{Pieters_2000}), Mercury (\cite{Lucey_2011}), and asteroids (\cite{BRITT_1994}). However, the particles in our models are significantly larger than the nanophase iron inclusions in lunar agglutinate rims--iron inclusions of this size ($>$25 nm) that tend to darken spectra without reddening \cite{LUCEY_2008, NOBLE_2007}. While submicroscopic particles are found throughout the space weathering literature, our Rayleigh scattering particles differ from those studied in the lunar context as they are not modeled as inclusions within a silicate matrix but as separate particles. Our Rayleigh scattering particles also may be different in composition, as our best-fitting models include both amorphous carbon and graphite as potential scattering species, along with the traditional iron. Additionally, studies of space weathering on asteroids focus mainly on C- and S- type asteroids, rather than the D- and P- type asteroids represented by Orus and Patroclus, respectively, so it is unclear how space weathering affects Trojans in particular. Further research is needed to distinguish between iron-based and carbon-based Rayleigh scatterers and understand the role these submicroscopic particles may play in space weathering on the Trojans.

\section{Conclusions}

Using the G280 grism on the Hubble Space Telescope, we measured near-ultraviolet and visible spectra of four Trojan asteroids in order to investigate potential absorption features in the near UV. Absorption features in the near-UV were expected based on previous ground and space-based observations of Trojans (e.g. \cite{Wong_2019, Karlson_2009, Zellner_1985}, as well as predicted compositional components of Trojan surfaces invoked to explain their infrared spectral features (e.g. \cite{Yang_2013, Wong_2016}). Though we did not observe the predicted absorption features, we detected a novel spectral feature: an increase in spectral reflectance from 0.35-0.3 $\mu$m. This ``bump" feature is present in the spectra of both Orus and the Patroclus-Menoetius binary. We developed a modified Hapke model in order to test whether this feature could be due to Rayleigh scattering from fine grained, submicroscopic particles on the surfaces of Trojans, testing two general classes of Rayleigh scatterers: ices (including H$_2$O, CO$_2$, and NH$_3$); and opaques (including iron, graphite, and amorphous carbon). We find that our spectra are consistent with a silicate surface mixed with opaques, including opaque Rayleigh scattering species. Though our results identify the opaque class of scatterers as a potential species responsible for producing the ``bump" feature, our model does not distinguish between the individual opaques species (iron, graphites, and amorphous carbon). We find that the feature could be caused by Rayleigh scattering particles ranging in size from 20 - 80 nm. Particles of this size have been implicated as space weathering agents in the literature, acting to darken the spectra of airless bodies (e.g. \cite{Lucey_2011, LUCEY_2008,Pieters_2000, BRITT_1994}). Whether these particles are the products of space weathering and responsible for the dark surfaces of the Trojan asteroids is beyond the scope of the research presented in this paper, but may be addressed by the upcoming Lucy mission, which will have the capability to measure the photometric properties of Trojan surfaces of different ages (\cite{Levison_2021}).

\section{Acknowledgements}

The authors wish to acknowledge Keith Noll for his assistance in preparing the Hubble Space Telescope proposal. We would also like to thank the two anonymous peer reviewers whose comments improved the quality and clarity of the manuscript. 

These observations are associated with programs \#15259 and \#15504. Support for programs \#15259 and \#15504 was provided by NASA through a grant from the Space Telescope Science Institute, which is operated by the Association of Universities for Research in Astronomy, Inc., under NASA
contract NAS5-26555.

These results made use of the Lowell Discovery Telescope (LDT) at Lowell Observatory.  Lowell is a private, non-profit institution dedicated to astrophysical research and public appreciation of astronomy and operates the LDT in partnership with Boston University, the University of Maryland, the University of Toledo, Northern Arizona University and Yale University. The upgrade of the DeVeny optical spectrograph has been funded by a generous grant from John and Ginger Giovale and by a grant from the Mt. Cuba Astronomical Foundation.

\bibliography{uvtrojans}{}

\begin{thebibliography}{}
\expandafter\ifx\csname natexlab\endcsname\relax\def\natexlab#1{#1}\fi
\providecommand{\url}[1]{\href{#1}{#1}}
\providecommand{\dodoi}[1]{doi:~\href{http://doi.org/#1}{\nolinkurl{#1}}}
\providecommand{\doeprint}[1]{\href{http://ascl.net/#1}{\nolinkurl{http://ascl.net/#1}}}
\providecommand{\doarXiv}[1]{\href{https://arxiv.org/abs/#1}{\nolinkurl{https://arxiv.org/abs/#1}}}

\bibitem[{Bendjoya {et~al.}(2004)Bendjoya, Cellino, Di~Martino, \&
  Saba}]{bendjoya-2004}
Bendjoya, P., Cellino, A., Di~Martino, M., \& Saba, L. 2004, Icarus, 168, 374,
  \dodoi{https://doi.org/10.1016/j.icarus.2003.12.004}

\bibitem[{Bertin \& Arnouts(1996)}]{Bertin_1996}
Bertin, E., \& Arnouts, S. 1996, Astronomy and astrophysics supplement series,
  117, 393, \dodoi{10.1051/aas:1996164}

\bibitem[{{Bida} {et~al.}(2014){Bida}, {Dunham}, {Massey}, \& {Roe}}]{DeVeny}
{Bida}, T.~A., {Dunham}, E.~W., {Massey}, P., \& {Roe}, H.~G. 2014, in Society
  of Photo-Optical Instrumentation Engineers (SPIE) Conference Series, Vol.
  9147, Ground-based and Airborne Instrumentation for Astronomy V, ed. S.~K.
  {Ramsay}, I.~S. {McLean}, \& H.~{Takami}, 91472N, \dodoi{10.1117/12.2056872}

\bibitem[{Binzel \& Sauter(1992)}]{BINZEL1992}
Binzel, R.~P., \& Sauter, L.~M. 1992, Icarus, 95, 222,
  \dodoi{https://doi.org/10.1016/0019-1035(92)90039-A}

\bibitem[{Britt \& Pieters(1994)}]{BRITT_1994}
Britt, D., \& Pieters, C. 1994, Geochimica et Cosmochimica Acta, 58, 3905,
  \dodoi{https://doi.org/10.1016/0016-7037(94)90370-0}

\bibitem[{Brown(2014)}]{BROWN_2014}
Brown, A.~J. 2014, Icarus, 239, 85,
  \dodoi{https://doi.org/10.1016/j.icarus.2014.05.042}

\bibitem[{Brown(2016)}]{Brown_2016}
Brown, M.~E. 2016, The Astronomical Journal, 152, 159,
  \dodoi{https://doi.org/10.3847/0004-6256/152/6/159}

\bibitem[{Buie {et~al.}(2021)Buie, Keeney, Strauss, Blank, Moore, Porter,
  Wasserman, Weryk, Levison, Olkin, {et~al.}}]{buie2021}
Buie, M.~W., Keeney, B.~A., Strauss, R.~H., {et~al.} 2021, The Planetary
  Science Journal, 2, 202, \dodoi{https://doi.org/10.3847/PSJ/ac1f9b}

\bibitem[{Cahill {et~al.}(2012)Cahill, Blewett, Nguyen, Xu, Kirillov, Lawrence,
  Denevi, \& Coman}]{Cahill_2012}
Cahill, J. T.~S., Blewett, D.~T., Nguyen, N.~V., {et~al.} 2012, Geophysical
  Research Letters, 39, \dodoi{https://doi.org/10.1029/2012GL051630}

\bibitem[{{Cheng} \& {WFC3 Team}(1999)}]{WFC3}
{Cheng}, E.~S., \& {WFC3 Team}. 1999, in American Astronomical Society Meeting
  Abstracts, Vol. 195, American Astronomical Society Meeting Abstracts, 85.06

\bibitem[{Clark {et~al.}(2007)Clark, Swayze, Wise, Livo, Hoefen, Kokaly, \&
  Sutley}]{clark_library_2007}
Clark, R.~N., Swayze, G.~A., Wise, R.~A., {et~al.} 2007, USGS digital spectral
  library splib06a, Tech. rep., US Geological Survey

\bibitem[{Clark {et~al.}(2008)Clark, Curchin, Jaumann, Cruikshank, Brown,
  Hoefen, Stephan, Moore, Buratti, Baines, {et~al.}}]{clark_2008}
Clark, R.~N., Curchin, J.~M., Jaumann, R., {et~al.} 2008, Icarus, 193, 372,
  \dodoi{https://doi.org/10.1016/j.icarus.2007.08.035}

\bibitem[{Clark {et~al.}(2012)Clark, Cruikshank, Jaumann, Brown, Stephan,
  Dalle~Ore, Livo, Pearson, Curchin, Hoefen, {et~al.}}]{clark_2012}
Clark, R.~N., Cruikshank, D.~P., Jaumann, R., {et~al.} 2012, Icarus, 218, 831,
  \dodoi{https://doi.org/10.1016/j.icarus.2012.01.008}

\bibitem[{Cloutis {et~al.}(2008)Cloutis, McCormack, Bell, Hendrix, Bailey,
  Craig, Mertzman, Robinson, \& Riner}]{Cloutis_2008}
Cloutis, E.~A., McCormack, K.~A., Bell, J.~F., {et~al.} 2008, Icarus, 197, 321,
  \dodoi{https://doi.org/10.1016/j.icarus.2008.04.018}

\bibitem[{Cruikshank {et~al.}(2001)Cruikshank, Dalle~Ore, Roush, Geballe, Owen,
  de~Bergh, Cash, \& Hartmann}]{cruikshank2001}
Cruikshank, D.~P., Dalle~Ore, C.~M., Roush, T.~L., {et~al.} 2001, Icarus, 153,
  348, \dodoi{https://doi.org/10.1006/icar.2001.6703}

\bibitem[{DeMeo \& Carry(2014)}]{DemeoCarry_2014}
DeMeo, F.~E., \& Carry, B. 2014, Nature, 505, 629,
  \dodoi{https://doi.org/10.1038/nature12908}

\bibitem[{{Devogele} \& {Moskovitz}(2019)}]{Devogele_SP}
{Devogele}, M., \& {Moskovitz}, N. 2019, in EPSC-DPS Joint Meeting 2019, Vol.
  2019, EPSC--DPS2019--841

\bibitem[{Dorschner {et~al.}(1995)Dorschner, Begemann, Henning, Jaeger, \&
  Mutschke}]{Dorschner_1995}
Dorschner, J., Begemann, B., Henning, T., Jaeger, C., \& Mutschke, H. 1995,
  Astronomy and Astrophysics, 300, 503

\bibitem[{Draine(1985)}]{Draine_1985}
Draine, B. 1985, The Astrophysical Journal Supplement Series, 57, 587

\bibitem[{Dressel(2021)}]{WFC3_handbook}
Dressel, L. 2021, Wide Field Camera 3 Instrument Handbook, Version 13.0

\bibitem[{Emery \& Brown(2004)}]{Emery_2004}
Emery, J., \& Brown, R. 2004, Icarus, 170, 131,
  \dodoi{https://doi.org/10.1016/j.icarus.2004.02.004}

\bibitem[{Emery {et~al.}(2006)Emery, Cruikshank, \& Van~Cleve}]{Emery_2006}
Emery, J., Cruikshank, D., \& Van~Cleve, J. 2006, Icarus, 182, 496,
  \dodoi{https://doi.org/10.1016/j.icarus.2006.01.011}

\bibitem[{Emery {et~al.}(2011)Emery, Burr, \& Cruikshank}]{Emery_2010}
Emery, J.~P., Burr, D.~M., \& Cruikshank, D.~P. 2011, The Astronomical Journal,
  141, 25, \dodoi{https://doi.org/10.1088/0004-6256/141/1/25}

\bibitem[{Fayolle {et~al.}(2021)Fayolle, Quirico, Schmitt, Jovanovic, Gautier,
  Carrasco, Grundy, Vuitton, Poch, Protopapa, Young, Cruikshank, {Dalle Ore},
  Bertrand, \& Stern}]{Fayolle_2021}
Fayolle, M., Quirico, E., Schmitt, B., {et~al.} 2021, Icarus, 367, 114574,
  \dodoi{https://doi.org/10.1016/j.icarus.2021.114574}

\bibitem[{Fornasier {et~al.}(2007)Fornasier, Dotto, Hainaut, Marzari,
  Boehnhardt, {De Luise}, \& Barucci}]{Fornasier_2007}
Fornasier, S., Dotto, E., Hainaut, O., {et~al.} 2007, Icarus, 190, 622,
  \dodoi{https://doi.org/10.1016/j.icarus.2007.03.033}

\bibitem[{Gartrelle {et~al.}(2021)Gartrelle, Hardersen, Izawa, \&
  Nowinski}]{gartrelle_2021}
Gartrelle, G.~M., Hardersen, P.~S., Izawa, M.~R., \& Nowinski, M.~C. 2021,
  Icarus, 354, 114043, \dodoi{https://doi.org/10.1016/j.icarus.2020.114043}

\bibitem[{Grav {et~al.}(2012)Grav, Mainzer, Bauer, Masiero, \&
  Nugent}]{Grav_2012}
Grav, T., Mainzer, A.~K., Bauer, J.~M., Masiero, J.~R., \& Nugent, C.~R. 2012,
  The Astrophysical Journal, 759, 49,
  \dodoi{https://doi.org/10.1088/0004-637X/759/1/49}

\bibitem[{Grundy {et~al.}(2018)Grundy, Noll, Buie, \&
  Levison}]{grundy2018upcoming}
Grundy, W., Noll, K., Buie, M., \& Levison, H. 2018, Icarus, 305, 198,
  \dodoi{https://doi.org/10.1016/j.icarus.2018.01.009}

\bibitem[{Guilbert-Lepoutre(2014)}]{GUILBERTLEPOUTRE2014}
Guilbert-Lepoutre, A. 2014, Icarus, 231, 232,
  \dodoi{https://doi.org/10.1016/j.icarus.2013.12.014}

\bibitem[{Hapke(1981)}]{Hapke_1981}
Hapke, B. 1981, Journal of Geophysical Research: Solid Earth, 86, 3039

\bibitem[{Hapke(1993)}]{Hapke_1993}
---. 1993, Theory of Reflectance Spectroscopy (Topics in Remote Sensing, vol.
  3),  Cambridge University Press, Cambridge

\bibitem[{Hapke {et~al.}(1981)Hapke, Wells, Wagner, \&
  Partlow}]{Hapke_1981_frost}
Hapke, B., Wells, E., Wagner, J., \& Partlow, W. 1981, Icarus, 47, 361

\bibitem[{Hendrix {et~al.}(2016)Hendrix, Vilas, \& Li}]{Hendrix_2016}
Hendrix, A.~R., Vilas, F., \& Li, J.-Y. 2016, Meteoritics \& Planetary Science,
  51, 105, \dodoi{https://doi.org/10.1111/maps.12575}

\bibitem[{Karlsson {et~al.}(2009)Karlsson, Lagerkvist, \&
  Davidsson}]{Karlson_2009}
Karlsson, O., Lagerkvist, C.-I., \& Davidsson, B. 2009, Icarus, 199, 106,
  \dodoi{https://doi.org/10.1016/j.icarus.2008.08.012}

\bibitem[{Kelley {et~al.}(2017)Kelley, Woodward, Gehrz, Reach, \&
  Harker}]{KELLEY2017}
Kelley, M.~S., Woodward, C.~E., Gehrz, R.~D., Reach, W.~T., \& Harker, D.~E.
  2017, Icarus, 284, 344, \dodoi{https://doi.org/10.1016/j.icarus.2016.11.029}

\bibitem[{Khare {et~al.}(1993)Khare, Thompson, Cheng, Chyba, Sagan, Arakawa,
  Meisse, \& Tuminello}]{Khare_1993}
Khare, B., Thompson, W., Cheng, L., {et~al.} 1993, Icarus, 103, 290

\bibitem[{Khare {et~al.}(1984)Khare, Sagan, Arakawa, Suits, Callcott, \&
  Williams}]{Khare_1984}
Khare, B.~N., Sagan, C., Arakawa, E., {et~al.} 1984, Icarus, 60, 127

\bibitem[{K{\"u}mmel {et~al.}(2009)K{\"u}mmel, Walsh, Pirzkal, Kuntschner, \&
  Pasquali}]{Kummel_2009}
K{\"u}mmel, M., Walsh, J., Pirzkal, N., Kuntschner, H., \& Pasquali, A. 2009,
  Publications of the Astronomical Society of the Pacific, 121, 59,
  \dodoi{https://doi.org/10.1086/596715}

\bibitem[{Lazzarin {et~al.}(1995)Lazzarin, Barbieri, \&
  Barucci}]{Lazzarin_1995}
Lazzarin, M., Barbieri, C., \& Barucci, M. 1995, The Astronomical Journal, 110,
  3058

\bibitem[{Levison {et~al.}(2021)Levison, Olkin, Noll, Marchi, III, Bierhaus,
  Binzel, Bottke, Britt, Brown, Buie, Christensen, Emery, Grundy, Hamilton,
  Howett, Mottola, Pätzold, Reuter, Spencer, Statler, Stern, Sunshine, Weaver,
  \& Wong}]{Levison_2021}
Levison, H.~F., Olkin, C.~B., Noll, K.~S., {et~al.} 2021, The Planetary Science
  Journal, 2, 171, \dodoi{10.3847/psj/abf840}

\bibitem[{Lucey \& Noble(2008)}]{LUCEY_2008}
Lucey, P.~G., \& Noble, S.~K. 2008, Icarus, 197, 348,
  \dodoi{https://doi.org/10.1016/j.icarus.2008.05.008}

\bibitem[{Lucey \& Riner(2011)}]{Lucey_2011}
Lucey, P.~G., \& Riner, M.~A. 2011, Icarus, 212, 451,
  \dodoi{https://doi.org/10.1016/j.icarus.2011.01.022}

\bibitem[{Luna {et~al.}(2012)Luna, Satorre, Domingo, Mill{\'a}n, \&
  Santonja}]{Luna_2012}
Luna, R., Satorre, M., Domingo, M., Mill{\'a}n, C., \& Santonja, C. 2012,
  Icarus, 221, 186, \dodoi{https://doi.org/10.1016/j.icarus.2012.07.016}

\bibitem[{Mainzer {et~al.}(2019)Mainzer, Bauer, Cutri, Grav, Kramer, Masiero, ,
  Wright, \& Eds.}]{Mainzer_NEOWISE}
Mainzer, A., Bauer, J., Cutri, R., {et~al.} 2019, NEOWISE Diameters and Albedos
  V2.0,  NASA Planetary Data System, \dodoi{10.26033/18s3-2z54}

\bibitem[{Marsset {et~al.}(2020)Marsset, DeMeo, Binzel, Bus, Burbine, Burt,
  Moskovitz, Polishook, Rivkin, Slivan, {et~al.}}]{marsset2020}
Marsset, M., DeMeo, F.~E., Binzel, R.~P., {et~al.} 2020, The Astrophysical
  Journal Supplement Series, 247, 73,
  \dodoi{https://doi.org/10.3847/1538-4365/ab7b5}

\bibitem[{Martonchik {et~al.}(1984)Martonchik, Orton, \&
  Appleby}]{Martonchik_1984}
Martonchik, J.~V., Orton, G.~S., \& Appleby, J.~F. 1984, Applied optics, 23,
  541

\bibitem[{Marzari \& Scholl(1998)}]{Marzari_1998}
Marzari, F., \& Scholl, H. 1998, Icarus, 131, 41

\bibitem[{Meftah {et~al.}(2018)Meftah, Dam{\'e}, Bols{\'e}e, Hauchecorne,
  Pereira, Sluse, Cessateur, Irbah, Bureau, Weber, {et~al.}}]{Meftah_2018}
Meftah, M., Dam{\'e}, L., Bols{\'e}e, D., {et~al.} 2018, Astronomy \&
  Astrophysics, 611, A1, \dodoi{https://doi.org/10.1051/0004-6361/201731316}

\bibitem[{Morbidelli {et~al.}(2005)Morbidelli, Levison, Tsiganis, \&
  Gomes}]{Morbidelli_2005}
Morbidelli, A., Levison, H.~F., Tsiganis, K., \& Gomes, R. 2005, Nature, 435,
  462, \dodoi{https://doi.org/10.1038/nature03540}

\bibitem[{Mueller {et~al.}(2010)Mueller, Marchis, Emery, Harris, Mottola,
  Hestroffer, Berthier, \& Di~Martino}]{Mueller_2010}
Mueller, M., Marchis, F., Emery, J.~P., {et~al.} 2010, Icarus, 205, 505,
  \dodoi{https://doi.org/10.1016/j.icarus.2009.07.043}

\bibitem[{Nesvorn{\`y} {et~al.}(2013)Nesvorn{\`y}, Vokrouhlick{\`y}, \&
  Morbidelli}]{Nesvorny_2013}
Nesvorn{\`y}, D., Vokrouhlick{\`y}, D., \& Morbidelli, A. 2013, The
  Astrophysical Journal, 768, 45,
  \dodoi{https://doi.org/10.1088/0004-637X/768/1/45}

\bibitem[{Noble {et~al.}(2007)Noble, Pieters, \& Keller}]{NOBLE_2007}
Noble, S.~K., Pieters, C.~M., \& Keller, L.~P. 2007, Icarus, 192, 629,
  \dodoi{https://doi.org/10.1016/j.icarus.2007.07.021}

\bibitem[{Pieters {et~al.}(2000)Pieters, Taylor, Noble, Keller, Hapke, Morris,
  Allen, McKay, \& Wentworth}]{Pieters_2000}
Pieters, C.~M., Taylor, L.~A., Noble, S.~K., {et~al.} 2000, Meteoritics \&
  Planetary Science, 35, 1101,
  \dodoi{https://doi.org/10.1111/j.1945-5100.2000.tb01496.x}

\bibitem[{Pirani {et~al.}(2019)Pirani, Johansen, Bitsch, Mustill, \&
  Turrini}]{pirani_2019}
Pirani, S., Johansen, A., Bitsch, B., Mustill, A.~J., \& Turrini, D. 2019,
  Astronomy \& Astrophysics, 623, A169

\bibitem[{Roig {et~al.}(2008)Roig, Ribeiro, \& Gil-Hutton}]{roig-2008}
Roig, F., Ribeiro, A., \& Gil-Hutton, R. 2008, Astronomy \& Astrophysics, 483,
  911, \dodoi{https://doi.org/10.1051/0004-6361:20079177}

\bibitem[{Rouleau \& Martin(1991)}]{Rouleau_1991}
Rouleau, F., \& Martin, P. 1991, The Astrophysical Journal, 377, 526

\bibitem[{Rozehnal {et~al.}(2016)Rozehnal, Bro{\v{z}}, Nesvorn{\`y}, Durda,
  Walsh, Richardson, \& Asphaug}]{rozehnal2016}
Rozehnal, J., Bro{\v{z}}, M., Nesvorn{\`y}, D., {et~al.} 2016, Monthly Notices
  of the Royal Astronomical Society, 462, 2319,
  \dodoi{https://doi.org/10.1093/mnras/stw1719}

\bibitem[{Sharkey {et~al.}(2019)Sharkey, Reddy, Sanchez, Izawa, \&
  Emery}]{sharkey_2019}
Sharkey, B.~N., Reddy, V., Sanchez, J.~A., Izawa, M.~R., \& Emery, J.~P. 2019,
  The Astronomical Journal, 158, 204,
  \dodoi{https://doi.org/10.3847/1538-3881/ab46c0}

\bibitem[{Sheppard \& Trujillo(2006)}]{Sheppard_2006}
Sheppard, S.~S., \& Trujillo, C.~A. 2006, Science, 313, 511,
  \dodoi{10.1126/science.1127173}

\bibitem[{Souza-Feliciano {et~al.}(2020)Souza-Feliciano, {De Prá},
  Pinilla-Alonso, Alvarez-Candal, Fernández-Valenzuela, {De León}, Binzel,
  Arcoverde, Rondón, \& Evangelista}]{SouzaFeliciano_2020}
Souza-Feliciano, A., {De Prá}, M., Pinilla-Alonso, N., {et~al.} 2020, Icarus,
  338, 113463, \dodoi{https://doi.org/10.1016/j.icarus.2019.113463}

\bibitem[{Vinogradova(2015)}]{vinogradova2015}
Vinogradova, T. 2015, Monthly Notices of the Royal Astronomical Society, 454,
  2436, \dodoi{https://doi.org/10.1093/mnras/stv2115}

\bibitem[{Wakeford {et~al.}(2020)Wakeford, Sing, Stevenson, Lewis, Pirzkal,
  Wilson, Goyal, Kataria, Mikal-Evans, Nikolov, {et~al.}}]{wakeford_2020}
Wakeford, H., Sing, D., Stevenson, K., {et~al.} 2020, The Astronomical Journal,
  159, 204, \dodoi{https://doi.org/10.3847/1538-3881/ab7b78}

\bibitem[{Warren(1984)}]{Warren_1984}
Warren, S.~G. 1984, Applied optics, 23, 1206

\bibitem[{Warren(1986)}]{Warren_1986}
---. 1986, Applied Optics, 25, 2650

\bibitem[{Wenger {et~al.}(2000)Wenger, Ochsenbein, Egret, Dubois, Bonnarel,
  Borde, Genova, Jasniewicz, Lalo{\"e}, Lesteven, {et~al.}}]{simbad}
Wenger, M., Ochsenbein, F., Egret, D., {et~al.} 2000, Astronomy and
  Astrophysics Supplement Series, 143, 9,
  \dodoi{https://doi.org/10.1051/aas:2000332}

\bibitem[{Westley {et~al.}(1998)Westley, Baratta, \& Baragiola}]{Westley_1998}
Westley, M.~S., Baratta, G., \& Baragiola, R. 1998, The Journal of chemical
  physics, 108, 3321

\bibitem[{Wong \& Brown(2016)}]{Wong_2016}
Wong, I., \& Brown, M.~E. 2016, The Astronomical Journal, 152, 90,
  \dodoi{https://doi.org/10.3847/0004-6256/152/4/90}

\bibitem[{Wong {et~al.}(2019)Wong, Brown, Blacksberg, Ehlmann, \&
  Mahjoub}]{Wong_2019}
Wong, I., Brown, M.~E., Blacksberg, J., Ehlmann, B.~L., \& Mahjoub, A. 2019,
  The Astronomical Journal, 157, 161,
  \dodoi{https://doi.org/10.3847/1538-3881/ab0e00}

\bibitem[{Wong {et~al.}(2014)Wong, Brown, \& Emery}]{Wong_2014}
Wong, I., Brown, M.~E., \& Emery, J.~P. 2014, The Astronomical Journal, 148,
  112, \dodoi{https://doi.org/10.1088/0004-6256/148/6/112}

\bibitem[{Yang {et~al.}(2013)Yang, Lucey, \& Glotch}]{Yang_2013}
Yang, B., Lucey, P., \& Glotch, T. 2013, Icarus, 223, 359,
  \dodoi{https://doi.org/10.1016/j.icarus.2012.11.025}

\bibitem[{{Ye} {et~al.}(2016){Ye}, {Hui}, Brown, Campbell-Brown, Pokorný,
  Wiegert, \& {Gao}}]{YE2016}
{Ye}, Q.-Z., {Hui}, M.-T., Brown, P.~G., {et~al.} 2016, Icarus, 264, 48,
  \dodoi{https://doi.org/10.1016/j.icarus.2015.09.003}

\bibitem[{Zellner {et~al.}(1985)Zellner, Tholen, \& Tedesco}]{Zellner_1985}
Zellner, B., Tholen, D., \& Tedesco, E. 1985, Icarus, 61, 355,
  \dodoi{https://doi.org/10.1016/0019-1035(85)90133-2}

\end{thebibliography}
\bibliographystyle{aasjournal}

\end{document}